\documentclass[iop]{emulateapj}
\usepackage{apjfonts}
\usepackage{natbib}
\usepackage{graphicx}
\usepackage{times}
\usepackage{xspace} 
\usepackage{booktabs} 

\slugcomment{\it Submitted --- 2018 March 5; Resubmitted --- 2018 May 17}
\shorttitle{Physical Scenarios for Abrupt AGN Fading}
\shortauthors{Stern et al.}

\def\ie{{i.e.}}
\def\eg{{e.g.}}
\def\etal{{et al.}}
\def\wise{{\it WISE}}
\def\neowise{{\it NEOWISE}}

\def\spitzer{{\it Spitzer}}

\def\rosat{{\it ROSAT}}

\def\swift{{\it Swift}}

\def\qso{WISE~J1052+1519}

\def\deg{\ifmmode {^{\circ}}\else {$^\circ$}\fi}
\def\kms{\ifmmode {\rm\,km\,s^{-1}}\else
    ${\rm\,km\,s^{-1}}$\fi}
\def\ergcm2s{\ifmmode {\rm\,erg\,cm^{-2}\,s^{-1}}\else
    ${\rm\,erg\,cm^{-2}\,s^{-1}}$\fi}
\def\ergAcm2s{\ifmmode {\rm\,erg\,cm^{-2}\,s^{-1}\,\AA^{-1}}\else
    ${\rm\,erg\,cm^{-2}\,s^{-1}\,\AA^{-1}}$\fi}
\def\ergs{\ifmmode {\rm\,erg\,s^{-1}}\else
    ${\rm\,erg\,s^{-1}}$\fi}
\def\kmsMpc{\ifmmode {\rm\,km\,s^{-1}\,Mpc^{-1}}\else
    ${\rm\,km\,s^{-1}\,Mpc^{-1}}$\fi}

\def\spose#1{\hbox to 0pt{#1\hss}}
\def\simlt{\mathrel{\spose{\lower 3pt\hbox{$\mathchar"218$}}
     \raise 2.0pt\hbox{$\mathchar"13C$}}}
\def\simgt{\mathrel{\spose{\lower 3pt\hbox{$\mathchar"218$}}
     \raise 2.0pt\hbox{$\mathchar"13E$}}}

\def\plotfiddle#1#2#3#4#5#6#7{\centering \leavevmode
\vbox to#2{\rule{0pt}{#2}}
\includegraphics{#1}}

\begin{document}

\title{A Mid-IR Selected Changing-Look Quasar \\
and Physical Scenarios for Abrupt AGN Fading}

\author{Daniel~Stern\altaffilmark{1}, 
Barry~McKernan\altaffilmark{2,3,4},
Matthew~J.~Graham\altaffilmark{5}, 
K.~E.~S.~Ford\altaffilmark{2,3,4},
Nicholas~P.~Ross\altaffilmark{6},
Aaron~M.~Meisner\altaffilmark{7,8},
Roberto~J.~Assef\altaffilmark{9},
Mislav~Balokovi\'{c}\altaffilmark{5,10},	% P200 observer (17jan)
Murray~Brightman\altaffilmark{5},
Arjun~Dey\altaffilmark{11},
Andrew~Drake\altaffilmark{5},			% P200 observer (17feb)
S.~G.~Djorgovski\altaffilmark{5}, 
Peter~Eisenhardt\altaffilmark{1} \&
Hyunsung~D.~Jun\altaffilmark{12}
% Nikita~Kamraj\altaffilmark{5},		% P200 observer (17jan)
}

\altaffiltext{1}{Jet Propulsion Laboratory, California Institute
of Technology, 4800 Oak Grove Drive, Mail Stop 169-221, Pasadena,
CA 91109, USA [e-mail: daniel.k.stern@jpl.nasa.gov]}

\altaffiltext{2}{Department of Science, Borough of Manhattan Community
College, City University of New York, New York, NY 10007, USA}

\altaffiltext{3}{Department of Astrophysics, American Museum of
Natural History, New York, NY 10024, USA}

\altaffiltext{4}{Graduate Center, City University of New York, 365
5th Avenue, New York, NY 10016, USA}

\altaffiltext{5}{Cahill Center for Astronomy and Astrophysics,
California Institute of Technology, 1216 E.  California Blvd.,
Pasadena, CA 91125, USA}

\altaffiltext{6}{Institute for Astronomy, SUPA, University of
Edinburgh, Royal Observatory, Edinburgh EH9 3HJ, UK}

\altaffiltext{7}{Berkeley Center for Cosmological Physics, Berkeley,
CA 94720, USA}

\altaffiltext{8}{Lawrence Berkeley National Laboratory, Berkeley,
CA, 94720, USA}

\altaffiltext{9}{N\'ucleo de Astronom\'ia de la Facultad de
Ingenier\'ia y Ciencias, Universidad Diego Portales, Av. Ej\'ercito
Libertador 441, Santiago, Chile}

\altaffiltext{10}{Harvard-Smithsonian Center for Astrophysics, 60
Garden St., Cambridge, MA 02138, USA}

\altaffiltext{11}{National Optical Astronomy Observatory, 950 N.
Cherry Ave., Tucson, AZ 85719, USA}

\altaffiltext{12}{School of Physics, Korea Institute for Advanced
Study, 85 Hoegiro, Dongdaemun-gu, Seoul 02455, Korea}

\begin{abstract} 

We report a new changing-look quasar, WISE~J105203.55+151929.5 at
$z=0.303$, found by identifying highly mid-IR variable quasars in
the {\it WISE}/{\it NEOWISE} data stream.  Compared to multi-epoch
mid-IR photometry of a large sample of SDSS-confirmed quasars, \qso\
is an extreme photometric outlier, fading by more than a factor of
two at $3.4$ and $4.6 \mu$m since 2009.  \swift\ target-of-opportunity
observations in 2017 show even stronger fading in the soft X-rays
compared to the \rosat\ detection of this source in 1995, with at
least a factor of fifteen decrease.   We obtained second-epoch
spectroscopy with the Palomar telescope in 2017 which, when compared
with the 2006 archival SDSS spectrum, reveals that the broad H$\beta$
emission has vanished and that the quasar has become significantly
redder.  The two most likely interpretations for this dramatic
change are source fading or obscuration, where the latter is strongly
disfavored by the mid-IR data.  We discuss various physical scenarios
that could cause such changes in the quasar luminosity over this
timescale, and favor changes in the innermost regions of
the accretion disk that occur on the thermal and heating/cooling
front timescales.  We discuss possible physical triggers that could
cause these changes, and predict the multiwavelength signatures
that could distinguish these physical scenarios.

\end{abstract}


\section{Introduction}

While variability has long been recognized as a distinguishing
feature of quasars \citep[\eg,][]{Matthews:63}, it has only been
in recent years that new generations of wide-area, multi-epoch
optical surveys have allowed systematic study of the extremes of
such behavior.  In addition to observing such rare phenomena as
periodic quasars \citep[\eg,][]{Graham:15, Graham:15b, Liu:15},
flaring quasars \citep[\eg,][]{Lawrence:16, Graham:17, Kankare:17},
extreme broad absorption line variability \citep[\eg,][]{Rafiee:16,
Stern:17}, and tidal disruption events \citep[\eg,][]{Arcavi:14,
Blagorodnova:17}, this work has also identified a new class of
``changing-look quasars'' in which the strong UV continuum and broad
hydrogen emission lines associated with unobscured quasars either
appear or disappear on timescales of years \citep[\eg,][]{LaMassa:15,
Macleod:16, Ruan:16a, Ruan:16b, Runnoe:16, Gezari:17, Yang:17}.
The physical processes responsible for these changing-look quasars
are still debated, but physical changes in the accretion disk
structure appear the more likely cause rather than changes in
obscuration.  These disk structural changes are presumed associated
with changes in black hole accretion rate.

Related changing-look phenomena have been seen in Seyfert galaxies
at lower luminosities for several decades, generally from multi-epoch
targeted studies of specific sources at either X-ray or optical
wavelengths \citep[\eg,][and references therein]{Tohline:76,
Goodrich:89, StorchiBergmann:95b, Shappee:14}.  Indeed, the term
``changing-look'' was initially used to describe sources whose X-ray
spectra changed appearance on timescales of years, switching from
reflection-dominated to Compton-thin, or vice versa
\citep[\eg,][]{Matt:03}.  In some cases, such as the nearby Seyfert
galaxies NGC~1365 and IC~751, the extreme X-ray variability is
clearly associated with rapid changes in the nuclear obscuration
\citep[\eg,][]{Risaliti:02, Walton:14, Rivers:15, Ricci:16}.  On
the other hand, \citet{Matt:03} argue that extreme X-ray spectral
changes are more typically associated with a temporary switching-off
of the nuclear radiation. As one example, \citet{McElroy:16} and
\citet{Husemann:16} discuss the case of Mrk~1018 which, over the
past four decades, has evolved from a Seyfert~1.9 galaxy to a
Seyfert~1 galaxy and then back to a Seyfert~1.9 galaxy.  Due to the
lack of associated changes in either the Balmer decrement or neutral
hydrogen absorbing column of this source, these companion papers
argue that intrinsic changes in the accretion disk flux rather than
variable extinction likely drove the spectral evolution of Mrk~1018.

Mid-IR monitoring provides a powerful new tool for both finding
changing-look quasars, and for probing the physical processes
responsible for the observed changes.  X-rays and UV continuum
emission from quasars come from regions extremely close to the
central supermassive black hole, with separations of a few to a few
tens of gravitational radii, $r_g \equiv G M_{\rm BH} / c^2$.  This corresponds
to distances of less than a light-day for typical quasars, and, as
long discussed in the context of the unified model of active
galactic nuclei (AGN), the sightline of the observer to this compact
region strongly impacts the observed appearance of a quasar at UV
and higher energies \citep[\eg,][]{Urry:95}.  In contrast, the mid-IR
emission of quasars predominantly comes from a dusty region beyond
the dust sublimation radius, implying parsec-scale distances.  Since
this larger-scale material, generally believed to be toroidal in
structure, is re-processing emission from the active nucleus, it
is both less sensitive to the observer's exact sightline
\citep[\eg,][]{Stern:05, Stern:12, Assef:13}, and is subject to a
substantial time delay relative to luminosity changes in the nuclear
regions \citep[\eg,][]{Jun:15b, Ichikawa:17}.

The {\it Wide-field Infrared Survey Explorer} \citep[{\it
WISE};][]{Wright:10} mission and its continuation as the {\it
Near-Earth Object WISE Reactivation} \citep[{\it NEOWISE};][]{Mainzer:14}
mission provide ideal data for identifying mid-IR selected changing-look
quasars.  Since 2010 January, the polar-orbit \wise\ satellite has
imaged the full sky approximately every six months.  \citet{Assef:18}
presents a catalog of \wise-selected AGN across most of the
extragalactic sky, with 4.5 million AGN candidates identified at
90\% reliability, and nearly 21 million AGN candidates identified
at 75\% completeness (but 51\% reliability).  As part of that work,
\citet{Assef:18} discuss the subset of 687 high-reliability AGN
identified as highly mid-IR variable during the first year of \wise\
observations.  Considering the subset of these sources not detected
at radio energies, so as to avoid blazars, they present one quasar,
WISEA~J142846.71+172353.1, whose broad H$\alpha$ emission has
disappeared between an SDSS spectrum obtained in 2008 and a Palomar
spectrum obtained in 2017.  Assef \etal\ (in prep.) discusses another
extreme mid-IR variable quasar identified from the first year of
\wise\ data.  \citet{Sheng:17} considers ten published changing-look
AGN and investigates their mid-IR lightcurves in the \wise\ and
\neowise\ data.  They find strong ($> 0.4$~mag) variability in all
ten cases, and they find the mid-IR variability to be consistent
with echoing the optical variability with the time lag expected for
dust reprocessing.  \citet{Sheng:17} argue that this result is
inconsistent with varying obscuration causing the changing-look
phenomenon, and they instead favor a scenario with variable AGN
accretion rates causing the photometric variability.

Here and in a companion paper, \citet{Ross:18}, we present the first
changing-look quasars identified from the combined \wise\ and
\neowise\ data streams.  This provides a longer selection baseline
than the sample of sources discussed in \citet{Assef:18}, which is
the only other published example of a mid-IR selected changing-look
quasar.  Our paper is organized as follows:  \S~2 presents the
selection of WISE~J105203.55+151929.5 (hereafter \qso), which is
the focus of this paper; \S~3 presents follow-up spectroscopic
observations at optical and X-ray energies, demonstrating the extreme
changes in this source; \S~4 presents a detailed discussion of the
possible physics that could explain abrupt fading (or brightening)
of a quasar; and we summarize our conclusions in \S~5.

Throughout this paper, we use AB magnitudes unless otherwise indicated
and we adopt the concordance cosmology, $\Omega_{\rm M} = 0.3$,
$\Omega_\Lambda = 0.7$ and $H_0 = 70\, \kmsMpc$.

% While variability was recognized as a common feature of quasars
% from shortly after their initial discovery \citep{Matthews:63}, it
% has only been in recent years that wide-area, multi-epoch surveys
% have begun to systematically study and quantify such behavior
% \citep[\eg,][]{Macleod:12, Graham:14}.  Such work has identified
% rare classes of quasars with unusual light-curves, such as periodic
% quasars \citep[\eg,][]{Graham:15, Graham:15b, Jun:15b, Liu:15},
% flaring quasars \citep[e.g.,][]{Drake:11, Lawrence:16, Graham:17},
% and quasars undergoing significant step-like changes in their
% photometric properties \citep[e.g.,][]{LaMassa:15, Gezari:17,
% Stern:17}.

% [{\it Daniel:  Summarize recent literature on changing-look quasars
% \citep[\eg][]{LaMassa:15, Macleod:16, Gezari:17, Stern:17} --- and
% mention NuSTAR-identified changing-look quasar buried in
% \citet{Alexander:13}, as well as curmudgeonly discussion of
% decades-long knowledge of such events, from X-rays
% \citep[\eg][]{Risaliti:02}, in Seyferts (REF), and 3C303 (Arjun -
% REF?). Briefly describe WISE/NEOWISE, and emphasize that this is
% the first changing-look quasar selected from mid-IR data, which
% provides an additional, powerful tool for distinguishing between a
% fading AGN and increased obscuration of the central engine.}]

% \section{Observations}

\section{Selection of \qso}

We extracted mid-IR $W1$ (3.4~$\mu$m) and $W2$ (4.6~$\mu$m) lightcurves
for $\sim$200,000 Sloan Digital Sky Survey (SDSS) spectroscopic
quasars from Data Release 3 (DR3) of the Dark Energy Camera Legacy
Survey (DECaLS\footnote{\url{legacysurvey.org/decamls/}}).  These
lightcurves span the period from the beginning of the \wise\ mission
in 2010 January through 2014 December, corresponding to the first-year
of {\it NEOWISE} operations.  Note that there is a gap in the
\wise\ data between 2011 February and 2013 September when the
satellite was in hibernation.  For most celestial locations, the
90-min orbit of \wise\ provides $\approx 12$ observations of a source
over a $\approx$1 day period, and a given celestial location is observed
every six months.  For this study, we combine the shorter-cadence
data, and we call each longer cadence co-added observation a single
``epoch'' of observations.  This means that we typically have four
epochs of photometry available, with separations ranging from six
months to a maximum of nearly five years.  The $W1$/$W2$ lightcurves
were obtained by performing forced photometry at the locations of
DECam-detected optical sources on un{\it WISE} epochal co-adds
\citep{Lang:14, Lang:16}.  While this approach means that we cannot
probe variability on timescales less than 1~day, the co-adds allow
photometry 1.4~mag deeper than the individual exposures and remove
virtually all single-exposure artifacts (\eg, cosmic rays and
satellites).

%
% FIGURE 1 - WISE QSO VARIABILITY HISTOGRAM
\begin{figure}
\plotfiddle{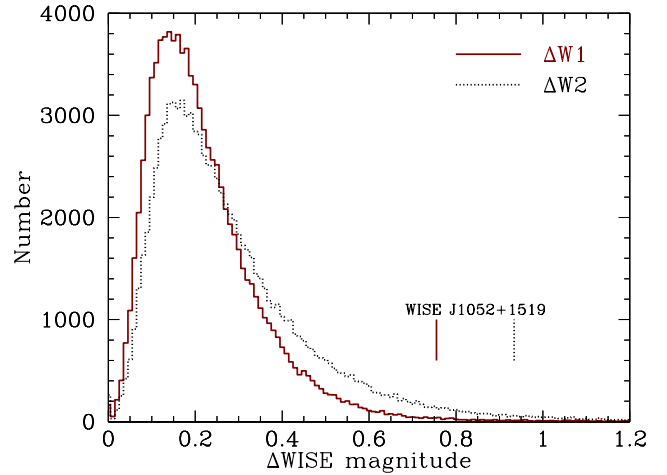}{2.45in}{-90}{32}{32}{-120}{190}
\caption{Histogram of maximum variation in mid-IR magnitude between
2010 January and 2014 December for 200,622 quasars in DECaLS DR3
region.  We only plot sources with mean epochal signal-to-noise
ratio $> 10$ for each band.  \qso\ stands out as being highly
variable in both $W1$ and $W2$.}
\label{fig:wisevar}
\end{figure}

Approximately 30,000 of the SDSS quasars with such $W1$/$W2$
lightcurves available are ``IR-bright'', in the sense that they are
above both the $W1$ and $W2$ single exposure thresholds and therefore
detected at very high significance in our co-adds. For this ensemble
of objects, the typical variation in each quasar's measured ($\vert
W1-W2 \vert$) color is 0.06 magnitudes, which includes statistical
and systematic errors.  Figure~\ref{fig:wisevar} presents the maximum
observed mid-IR variability of quasars over the five years of
observations, considering only those quasars detected with mean
signal-to-noise ratio $> 10$ in the individual epochs.  Most quasars
vary by less than 0.2~mag, with a small fraction varying by more
than a factor of two in flux over this 4-year period (i.e., $\Delta
m \ge 0.75$~mag).

% A full characterization of the typical mid-IR quasar variability
% will be presented separately (Ross \etal, in preparation).

% The typical measured single-band scatter is 0.07 magnitudes in each
% of $W1$ and $W2$.

% \bigskip
% [{\it Note from Arjun:  Can we be more quantitative, since much
% depends on the luminosity and color and redshift of the QSOs in the
% SDSS sample.  Perhaps this is for Aaron's sample description paper
% (which we should reference here), but we could at least mention the
% range of W1-W2 and perhaps r-W1 of the detected sample.}]
% \bigskip

To identify the most extreme outliers relative to these trends, we
selected objects with the following characteristics:

\begin{itemize}

\item monotonic variation in both $W1$ and $W2$;

\item $W1$ versus $W2$ flux correlation coefficient $>$0.9; and

\item $>0.5$ mag peak-to-peak variation in either $W1$ or $W2$.

\end{itemize}

\noindent This yielded a sample of 248 sources, of which 31 are
assumed to be blazars due to the presence of radio counterparts in
the Faint Images of the Radio Sky at Twenty cm survey
\citep[FIRST;][]{Becker:95}.  Another 22 are outside the FIRST
footprint, leaving 195 radio-undetected ($S_{\rm 1.4~GHz} \simlt
1$~mJy; $5 \sigma$) quasars in our IR-variable sample.  Note
that we did not impose our IR-bright criterion in selecting this
sample. Doing so in future explorations would further cull the
sample, potentially allowing us to loosen or remove the monotonicity
and correlation coefficient requirements and thereby capture a
richer variety of light curve behaviors.  We selected five of these
objects for follow-up spectroscopy with Palomar on the night of UT
2017 January 30. \qso, one of these five, had a peak-to-peak variation
of 0.76 (0.93) mags in $W1$ ($W2$) between 2010 May and 2014 December,
and thus became 0.15 mags bluer in ($W1-W2$). This made it a
significant outlier in both single-band and IR color variability.

%
% FIGURE 2 - LIGHTCURVES
\begin{figure}
\plotfiddle{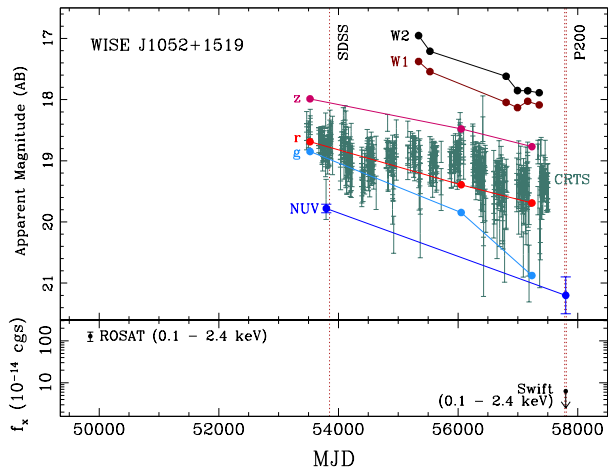}{2.6in}{-90}{32}{32}{-120}{190}
\caption{Multi-wavelength lightcurves of \qso, showing the epochs
of the spectroscopy with vertical dotted lines.  CRTS photometry
(green error bars) is unfiltered CCD observations \citep[for details,
see][]{Drake:09}.  In chronological order, the optical $grz$
photometry is from SDSS, Pan-STARRS, and DECaLS.   NUV photometry
is {\it GALEX} {\it NUV} ($\lambda_{\rm eff} = 2271$~\AA) for the
first epoch and {\it Swift}/UVOT UVM2 ($\lambda_{\rm eff} = 2231$~\AA)
for the second epoch.  The bottom panel shows the 0.1--2.4~keV X-ray
lightcurve, where the \rosat\ point corresponds to the detection
in 1995, and the \swift\ point corresponds to the 3$\sigma$ upper
limit for $\Gamma = 2$.  \qso\ has faded by more than a factor of
15 in the X-rays between the \rosat\ and \swift\ observations.}
% [{\it MATTHEW, AARON -- CAN WE ADD MORE RECENT CRTS AND UNWISE
% PHOTOMETRY?}]
\label{fig:lc}
\end{figure}

\subsection{UV to Mid-IR Photometry and Lightcurve}

Figure~\ref{fig:lc} presents a multi-wavelength lightcurve of \qso,
including optical data from SDSS, the Panoramic Survey Telescope
and Rapid Response System \citep[Pan-STARRS;][]{Chambers:16}, DECaLS,
and the Catalina Real-time Transient Survey \citep[CRTS;][]{Drake:09},
as well as the mid-IR photometry discussed above.  We have also
supplemented the $W1$/$W2$ lightcurves discussed above with {\it
NEOWISE} epochal co-adds between 2015 May and 2015 December,
providing two additional mid-IR photometric epochs to the lightcurve
in Figure~\ref{fig:lc}.

% Much of this data is presented in Table~\ref{table:photometry},
% which also includes the single-epoch UV photometry from the {\it
% Galaxy Evolution Explorer} \citep[{\it GALEX};][]{Martin:05}, and
% near-IR photometry from the Two Micron All Sky Survey
% \citep[][]{Skrutskie:06}.

\begin{figure}
\plotfiddle{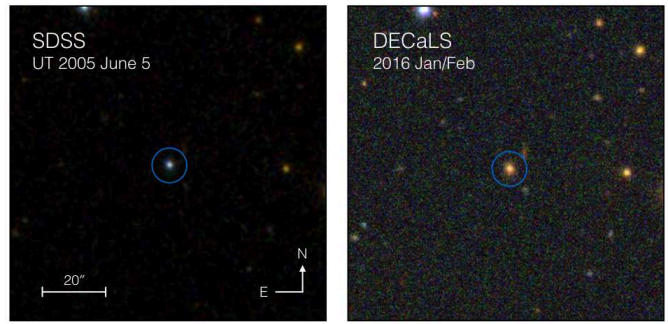}{2.0in}{90}{34}{34}{133}{-40}
\caption{False-color optical images of \qso\ from SDSS in 2005
(left) and DECaLS in 2016 (right).  In the earlier epoch image,
\qso\ has a color similar to the blue star to the NNE, while a
decade a later, the source is considerably redder, comparable in
color to the galaxies to the NW.}
\label{fig:image}
\end{figure}

Overall, \qso\ has faded significantly in the past decade at
both optical and mid-IR wavelengths.  Optically, the source has
changed most at the bluest wavelengths, fading by 1.8~mag at $g$-band
between SDSS and DeCaLS, though only by 1.0~mag at $r$-band and
0.8~mag at $z$-band.  Figure~\ref{fig:image} demonstrates this
chromatic change, showing side-by-side false-color optical images
from SDSS (2005) and DECaLS (2016):  compared to other sources in
the field, \qso\ has become significantly redder at optical wavelengths
over the past decade.  In the IR, the situation is reversed, with
the more significant fading occurring in the redder $W2$ band, by
0.9~mag, while the bluer $W1$ band faded by only 0.7~mag.  Since
the spectra of galaxies with ages above a few Myr peak in the
near-IR, while AGN spectra peak in the UV and mid-IR, with a dip
around $1 \mu$m \citep[\eg,][]{Assef:10}, AGN become most evident
in the rest-frame UV and long-wards of a few microns.  The observed
photometric evolution of \qso\ is therefore consistent with a
dramatic decrease in its AGN emission.  Notably, SDSS morphologically
classifies \qso\ as stellar, while DECaLS classified the source as
``D'', meaning a non-stellar source best fit with a de~Vaucouleurs
profile.

%
% FIGURE 3 - IMAGES

\section{Follow-Up Observations}

\subsection{Optical Spectroscopy}

\begin{figure}
\plotfiddle{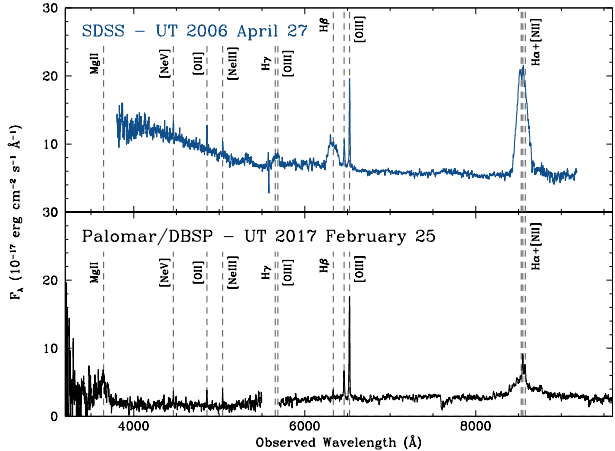}{2.6in}{-90}{32}{32}{-120}{190}
\caption{Multi-epoch optical spectra of \qso. Top panel shows the
SDSS spectrum from Spring 2006, while lower panel shows the Palomar
spectrum from Winter 2017.  Key emission lines are labelled.  In
the past decade, the continuum has become significantly less blue,
and higher order Balmer lines are no longer visible.}
\label{fig:spectra}
\end{figure}

\qso\ was first observed spectroscopically by SDSS on UT 2006 April
27 (MJD = 53852).  The spectrum, shown in Figure~\ref{fig:spectra},
reveals a typical quasar with broadened emission lines from multiple
hydrogen Balmer transitions (i.e., H$\alpha$ through H$\delta$),
strong, narrow emission from the [\ion{O}{2}] and [\ion{O}{3}]
doublets, and strong blue continuum rising below $\sim 5500$~\AA.
SDSS reports a redshift of $z = 0.303$ for the source, and classifies
it as a quasar.

Because of its unusual lightcurve, we obtained additional optical
spectroscopy of \qso\ using the Double Spectrograph (DBSP) on the
Hale 200'' Telescope at Palomar Observatory on UT 2017 January 30
(MJD = 57783).  We obtained a single 900~s observation of the target
using the 1\farcs5 slit at the parallactic angle.  The night was
photometric, though extremely windy, leading to highly variable
seeing that exceeded 3\arcsec\ (FWHM) at times.  The poor seeing
and image motion significantly compromised the quality of this
spectroscopic observation.  We processed the data using standard
procedures and flux calibrated the spectrum with observations of
the white dwarf spectrophotometric standard stars G191-B2B and HZ44
from \citet[][]{Massey:90} obtained on the same night.  Though with
limited signal-to-noise ratio, the January data showed a spectrum
with significantly less rest-frame UV/blue emission, and much weaker
Balmer emission, suggesting a dramatic spectral change since the
SDSS data from a decade earlier.

In order to improve the signal-to-noise ratio, we re-observed \qso\
with DBSP on the Palomar 200'' on UT 2017 February 25 (MJD = 57809).
We obtained two 600~s observations at the parallactic angle through
the 1\farcs5 slit in photometric, good-seeing conditions.  The
February data were flux calibrated with observations of the
spectrophotometric standards HZ14 and Feige~56 observed on that
same night.  Figure~\ref{fig:spectra} presents the February Palomar
data, which is of significantly better quality than the January
Palomar spectrum.

%
% FIGURE 4 - OPTICAL SPECTRA

Comparing the SDSS and Palomar spectra, we see that the source has
faded significantly over the past decade.  The rising blue continuum
is no longer evident, nor are any of the broad Balmer emission lines
other than H$\alpha$, which has faded by more than a factor of
three.  The H$\alpha$ line has also become broader as the
source faded, as expected.  Specifically, the line width increases
by a factor of $\sim 1.5-2$ (depending on whether single or double
Gaussian fits to the line are adopted) while the 5100~\AA\, continuum
fades by a factor of $\sim 2$; using the \citet{Jun:15a} black hole
mass estimator, we find consistent inferred black hole masses between
the SDSS and Palomar spectra. Broad \ion{Mg}{2} is visible in the
Palomar data, but is blue-ward of the SDSS spectral range and so
is not available for direct investigation of temporal evolution.
The narrow [\ion{O}{3}] lines are slightly weaker in the Palomar
data at the few tens of percent level.  Since the narrow-line region
of quasars are typically spatially extended on scales of tens to
thousands of parsecs, this difference is likely due to the larger,
3\arcsec\ diameter fiber used by SDSS compared to the 1\farcs5 wide
slit used at Palomar.

From the SDSS spectrum, \citet{Shen:11} measure the Balmer emission
lines to have full-widths at half-maximum of FWHM(H$\alpha) = 5113
\pm 129\, {\rm km}\, {\rm s}^{-1}$ and FWHM(H$\beta) = 5622 \pm
145\, {\rm km}\, {\rm s}^{-1}$, and the 5100~\AA\ continuum luminosity
to be $\log L(5100)/({\rm erg}\, {\rm s}^{-1}) = 44.107 \pm 0.004$.
\citet{Shen:11} also reports the H$\alpha$ line luminosity to be
$\log L({\rm H}\alpha)/({\rm erg}\, {\rm s}^{-1}) = 42.869 \pm
0.015$.  Using the \citet{Jun:15a} black hole mass estimator based
on FWHM(H$\alpha$) and $L(5100)$, we derive a black hole mass for
\qso\ of $\log M_{\rm BH}/M_\odot = 8.61 \pm 0.12$. The \citet{Jun:15a}
estimator based on FWHM(H$\alpha$) and $L$(H$\alpha$) implies a
comparable black hole mass of $\log M_{\rm BH}/M_\odot = 8.66 \pm
0.13$.  Adopting the bolometric correction of $BC_{5100} = 9.26$
from \citet{Shen:11}, this implies the quasar had an Eddington ratio
$\lambda_{\rm Edd}$ of 2\%\ in the SDSS spectrum.

\subsection{X-Ray Observations}

Fortuitously, \qso\ resides 40.6\arcmin\ from the spiral galaxy
MCG+03-28-022 ($z = 0.022$), which hosted the exceptionally luminous
type~II supernova SN~1988Z.  At the time, it was the most distant
and most luminous supernova detected at both radio \citep[][]{VanDyk:93}
and X-ray energies \citep[][]{Fabian:96}.  This is presumably due
to the supernova having exploded in a high-density environment,
potentially associated with mass loss from its high-mass progenitor
\citep[][]{Stathakis:91}.  \qso\ was serendipitously detected in
the 12.3~ks {\it ROSAT} follow-up X-ray observations of SN~1988Z
obtained in 1995~May, and is listed as 2RXS~J105203.9+151930 in the
Second {\it ROSAT} All-Sky Survey \citep[][]{Boller:16}. It is a
$\sim 4\sigma$ detection, with a 0.1--2.4~keV flux of $1.3 \times
10^{-12}\, {\rm erg}\, {\rm cm}^{-2}\, {\rm s}^{-1}$ (assuming a
power-law fit to the X-ray spectrum).

This relatively bright archival detection of our quasar prior to
its optical fading inspired a successful {\it Swift} Target of
Opportunity proposal to study how it has evolved at higher energies.
{\em Swift} observed \qso\ with the X-Ray Telescope
\citep[XRT;][]{Burrows:05} and Ultraviolet/Optical Telescope
\citep[UVOT;][]{Roming:05} instruments on UT~2017 February 17 (obsID
00034933001).  We processed the raw data using the online analysis
tools provided by the ASI Science Data
Center\footnote{\url{http://www.asdc.asi.it/mmia/index.php?mission=swiftmastr}}
using HEAsoft version 6.20 and CALDB version 20111031.  The total
exposure was 4.3~ks with {\it Swift}/XRT, and 3.3~ks with {\it
Swift}/UVOT (after co-adding five separate frames taken with the
UVM2 filter).

No X-ray source was detected at the coordinates of the {\it ROSAT}
counterpart of \qso. The nearest detected source (SNR=4.4) is $\approx
3.5\arcmin$ offset at 10:51:50.46, +15:17:53.26, with a 0.3--10~keV
count rate of $(6.0 \pm 1.4)\times 10^{-3}$\,s$^{-1}$.  Nearby
$2 \sigma$ detected sources have count rates $(1.3\pm0.7)\times
10^{-3}$\,s$^{-1}$ and $(1.6\pm0.8)\times10^{-3}$\,s$^{-1}$. We
therefore estimate a $3 \sigma$ upper limit on the count rate of
W1052+1519 to be $3\times 10^{-3}$\,s$^{-1}$ in the 0.3--10~keV
band. Assuming a simple unabsorbed power-law spectrum with a photon
index $\Gamma=2$, this corresponds to a flux limit of $1.07\times
10^{-13}$\,erg\,cm$^{-2}$\,s$^{-1}$ in the 0.3--10~keV band, and
$4.90\times10^{-14}$\,erg\,cm$^{-2}$\,s$^{-1}$ in the 2--10~keV
band. Alternatively, for a softer $\Gamma=2.5$ spectrum, the
corresponding limits are $8.89\times 10^{-14}$\,erg\,cm$^{-2}$\,s$^{-1}$
and $2.30\times 10^{-14}$\,erg\,cm$^{-2}$\,s$^{-1}$, respectively.
This conversion to flux includes the Galactic column density, $N_{\rm
H}=2.3\times 10^{20}$\,cm$^{-2}$ \citep{Kalberla:05}.  The X-ray
non-detection implies that \qso\ has faded by at least an order of
magnitude in the past two decades (Figure~\ref{fig:lc}).

The target is detected at low significance ($\lesssim 5$ sigma) in
the {\em Swift}/UVOT UVM2 image at coordinates 10:52:03.5, +15:19:28.9.
From aperture photometry using a 5-pixel (1\farcs8) radius, we
estimate the source magnitude to be $21.2\pm0.3$ in the UVM2 filter
($\lambda_{\rm eff}=2231$\,\AA). The background was determined from
an annulus with an inner radius of 15 and outer radius of 30 pixels,
including no other sources.

% Correcting for the reddening along the line of sight (E(B$-$V)$=$0.026;
% \citealt{Schlafly:11}), we find a flux density of $(3.12\pm0.81)\times
% 10^{-6}$\,Jy.

% NOTE: This is nufnu = 4.20e-14+/-1.09e-14 erg/s/cm2 at nu =
% 1.3450e+15 Hz.

% \section{Discussion}
% 
% \subsection{Black Hole Mass and Eddington Ratio}
% \subsection{SED Modeling}
% 
% [{\it Roberto: Multi-epoch SED modeling, which will presumably
% indicate fading over obscuration as the primary cause of the dramatic
% optical spectral changes. Think about related Figure(s).}]

\section{Physical Scenarios for Abrupt AGN Fading}
% {Explanations and Predictions}
\label{sec:explanations}

We next discuss possible physical explanations for the extreme
dimming observed for \qso, including testable predictions where
possible.  There are two broad categories of explanation for our
observations:  obscuration and changes to the inner disk.

\subsection{Obscuration}

\subsubsection{Obscuration by a Cloud in a Keplerian Orbit}

In principle, obscuration by a large cloud in a Keplerian orbit
passing along our line of sight to \qso\ provides a very simple and
natural explanation for the observed quasar dimming, similar to
what explains the strong (X-ray) variability of nearby changing-look
AGN such as NGC~1365 \citep[\eg,][]{Walton:14}.  This scenario,
however, has several clear predictions, providing the potential to
rule it out.

First, obscuration would imply that all wavelengths would dim
essentially simultaneously, with the depth of the dimming in each
waveband depending on the properties of the obscurer.  A well-sampled
multi-wavelength lightcurve could test this model with sufficient data.
Unfortunately, however, with only a handful of photometric points
over a ten-year period in several key wavebands, the multi-wavelength
lightcurves of \qso\ are too sparse to reliably determine whether
or not there are lags between the various wavebands.

Second, an obscuring medium should have a strong wavelength dependence,
with bluer photons being much more heavily extinguished than IR photons.
For a standard $R_V = 3.1$ \citet{Fitzpatrick:99} Milky Way reddening
law, the extinction in $U$-band (rest-frame 3500 \AA; observed
$\approx g$-band for our source) is approximately 25 times higher
than the extinction in $L$-band (rest-frame 3.5 $\mu$m; observed
$\approx W2$ for our source); \ie, $A_U/E(B-V) \sim 4.3$ compared
to $A_L/E(B-V) \sim 0.2$ \citep{Schlafly:11}.   This implies that
the observed 1.8~mag $g$-band fading of \qso\ should have been
accompanied by negligible ($< 0.1$~mag) fading in the mid-IR bands.
This is significantly less than what we see in Figure~\ref{fig:lc}.

Finally, as discussed in \citet{LaMassa:15} and \citet{Sheng:17},
even with conservative assumptions, the crossing time for a cloud
in a Keplerian orbit to occult the broad-line region would be a
decade to several decades.  This increases to well over a century
to occult the inner regions of the mid-IR emitting dusty torus,
which is the relevant emitting region for mid-IR selected changing-look
quasars.  This is substantially longer than the observed timescales,
providing further strong evidence against obscuration by a cloud
in a Keplerian orbit being the physical cause of the extreme
variability in changing-look quasars.

We also briefly note that a cloud obscuring just the inner regions
of the AGN accretion disk is an unlikely solution to explain the
mid-IR variability.  The mid-IR emission comes from a large,
parsec-scale region, likely toroidal in nature, which re-processes
higher energy photons coming from the compact central engine, 100's
of AU in scale.  A cloud in a Keplerian orbit can easily occult
that central region from the observers' line-of sight, and, indeed,
strong evidence of that has been seen at X-ray energies for some
systems.  Ignoring the challenges of the longevity of a cloud
surviving interior to the dust sublimation radius, to cause mid-IR
dimming, an implausibly large fraction of the sightlines between
the central regions and the torus would need to be obscured.
Furthermore, such an obscuring, large-scale cloud would be heated
in the process, and become a new source of mid-IR thermal emission,
thereby decreasing the level of observed mid-IR variability.  These
arguments all demonstrate that obscuration by a cloud in a Keplerian
orbit is an extremely unlikely as a physical scenario to explain
changing-look quasars with strong mid-IR variability.

\subsubsection{Obscuration by an Infalling Cloud}

In order to bypass some of the challenges in terms of timescale and
cloud size for obscuration by a dusty cloud in a Keplerian orbit,
some authors have suggested that an infalling obscuring cloud
provides a plausible alternative \citep[\eg,][]{Guo:16}.  However,
there are multiple challenges to this scenario.

First, clouds typically will not be infalling, but instead will be
in Keplerian orbits.  In order to be infalling, the clouds need to lose
their angular momentum.  Though this could potentially occur in a
collision with an approximately equal mass cloud of opposite angular
momentum, such a collision would destroy the clouds, or, at least,
leave something less coherent.  Assuming the clouds survives, the
freefall timescale $t_{\rm ff}$ for the cloud to infall from the
dust sublimation radius $R_{\rm dust}$ \citep[where we assume a
typical dust composed of silicate and graphite grains;][]{Mor:12},
\begin{eqnarray}
R_{\rm dust} \approx 0.4~{\rm pc}\, \left( L_{\rm bol} \over
{10^{45}\, {\rm erg}\, {\rm s}^{-1}} \right)^{1/2} \left( {1500\,
{\rm K}} \over T_{\rm sub} \right)^{2.6},
\end{eqnarray}
\noindent is given by
\begin{eqnarray}
t_{\rm ff} \sim 100~{\rm yr}\, \left( R \over {0.4~{\rm pc}} \right)^{3/2}
\left(M_{\rm BH} \over {10^8 M_\odot} \right)^{-1}.
\end{eqnarray}
\noindent For typical cloud sizes $R_{\rm cloud}$ and overdensities
$\delta_{\rm cloud} = \rho_{\rm cloud} / \rho_{\rm medium}$, this
is comparable to the cloud-crushing time $t_{\rm cc}$, i.e., the
timescale on which Kelvin-Helmholtz instabilities will shred the
cloud into a fragmented, comet-like structure,
\begin{eqnarray}
t_{\rm cc} \sim 100~{\rm yr}\, \left( \delta_{\rm cloud} \over 10^6
\right)^{1/2} \left( R_{\rm cloud} \over {4 \times 10^{10}\, {\rm
km}} \right) \left( v_{\rm rel} \over {10^4\, {\rm km}\, {\rm
s}^{-1}} \right)^{-1}
\end{eqnarray}
\noindent where $v_{\rm rel}$ is the velocity of the cloud relative
to the medium in which it is infalling \citep{Klein:94}.  Though
this calculation ignores the timescale on which the dusty, infalling,
fragmenting cloud would be sublimated, this scenario suggests that
an infalling, intact cloud would cause changes to the X-ray/UV
spectrum on a timescale of several decades, not the month-scale
dramatic variability observed in the changing-look quasar
SDSS~J231742.60+000535.1 discussed by \citet{Guo:16}.  Furthermore,
as noted earlier in the context of a Keplerian cloud obscuring the
innermost regions of an AGN from the observer, this infalling
scenario should not significantly affect the mid-IR luminosity of
the AGN.

The arguments above emphasize the added value of mid-IR selection
for identifying changing-look quasars.  Namely, the large physical
scale of the mid-IR emitting region combined with the longer wavelength
baseline provided by mid-IR data together impose strong constraints
for distinguishing intrinsic luminosity changes from obscuration.
Having shown that obscuration is an unlikely cause of the changing-look
quasars, particularly for mid-IR selected changing-look quasars
such as \qso, we next consider changes to the innermost regions of
the accretion disk as driving the observed fading.

% First we consider dimming of the central continuum due to obscuration
% along the sightline. On one hand, if there is no time-lag between
% the drop in the  optical/UV ('high energy') continuum and the
% reprocessed IR ('low energy') continuum, then the obscurer must lie
% at some distance from the reprocessing region. The size of the
% obscurer and the odds of an eclipse-event obscuring half of the
% low-energy region drop rapidly the further the obscurer is from the
% low-energy region. On the other hand, if there is a time-lag between
% the drop in the high-energy and low-energy continua, an obscuration
% model implies an obscurer lying between the high energy and low-energy
% regions.  Based on our observations, such an obscurer needs to hide
% $\sim 1/2$ of the high-energy continuum from the reprocessing region
% and $\sim$ all of the high-energy continuum from our sightline. A
% large-scale change in the nature of the inner disk could do this
% (see below), but a typical broad-line region cloud ($N_{H} \sim
% 10^{21-22}\rm{cm}^{-2}$, REF) will not reduce the 2-10keV X-ray
% continuum by an order of magnitude. We therefore think obscuration
% independent of inner disk changes is unlikely as a cause of the
% observed effects.

\subsection{Timescales for Changes to the Inner Disk}
\label{sec:times}

For possible explanations involving the accretion disk, we need
estimates of the relevant timescales for processes at small radii
\citep[\eg,][]{Czerny:06}.  Four timescales are important to consider
for our purposes:  the orbital, thermal, cooling/heating front, and
viscous timescales.

The orbital, or dynamical, timescale in the accretion disk is
approximately $t_{\rm orb} \sim 1/\Omega$ where $\Omega=\sqrt{GM_{\rm
BH}/R^{3}}$ is the Keplerian orbital angular frequency at radius
$R$.  This timescale is relevant for various processes, such
as the timescale on which the disk achieves hydrostatic equilibrium,
or the timescale on which magnetic loops with feet on the disk
surface become entangled.  The thermal timescale, corresponding
to the timescale on which the disk heats or cools, is $t_{\rm th}
\sim t_{\rm orb}/\alpha$, where $\alpha$ is the disk viscosity
parameter \citep{Shakura:73}. Cooling and heating fronts
propagate radially through an $\alpha$ disk at approximately the
sound speed multiplied by the viscosity pararameter, $\alpha$
\citep[\eg,][]{Hameury:09} --- \eg, in the limit of $\alpha = 1$,
the fronts propagate at the sound speed, while in less viscous disks
(\ie, with smaller $\alpha$), it becomes harder for the front to
propogate.  In the limit of $\alpha = 0$, there is no communication
between neighboring annuli in the disk, so the fronts cannot
propagate.  Indeed, in such a situation, there is no accretion
because there is no mechanism, like turbulence, to communicate
between neighboring disk annuli and dissipate angular momentum.
These heating and cooling fronts therefore cross the disk on
timescales of $t_{\rm front} \sim (h/R)^{-1} t_{\rm th}$, where
$h/R$ is the disk aspect ratio, with $h = c_{s}/\Omega$ the disk
height, and $c_{s}$ is the local disk sound speed at radius $R$.
Finally, the viscous disk timescale, the characteristic timescale
of mass flow, is $t_{\nu}=(h/R)^{-2}t_{\rm th}$.

Using the parameters for \qso\ derived at the end of \S 3.1, let
us assume a supermassive black hole with mass $M_{\rm BH} = 4 \times
10^{8}M_{\odot}$ and bolometric luminosity $L_{\rm bol} \sim 0.02\,
L_{\rm Edd}$, where $L_{\rm Edd}$ is the Eddington luminosity and
$\lambda_{\rm Edd}$ is the Eddington ratio, $\lambda_{\rm Edd} \equiv
L_{\rm bol} / L_{\rm Edd}$.  Then, the characteristic distance scale
is the gravitational radius, $r_g = GM_{\rm BH}/c^2 \sim 5.9 \times 10^{11}$m
($\sim 4$~AU).  Assuming $L =\eta \dot{M}c^{2}$ is the luminosity
due to accretion, where $\eta \sim 0.1$ is the luminosity efficiency
of accretion, the characteristic mass-flow rate is
\begin{equation}
\dot{M} \approx 0.2\, M_{\odot}\, {\rm yr}^{-1} \left(
\frac{\eta}{0.1}\right)^{-1} \left( \frac{M_{\rm BH}}{4 \times
10^{8}M_{\odot}}\right)\left(\frac{\lambda_{\rm Edd}}{0.02}\right).
\label{eq:m_dot}
\end{equation}
\noindent Assuming a standard (relatively) thin-disk AGN model
consisting of multiple annuli at temperatures that drop with radius,
the mass flow rate across each annulus is $\dot{M} = 3\pi\nu\Sigma$,
where $\nu$ is the viscosity, $\Sigma$ is the disk surface density,
and the disk temperature drops as $T \propto R^{-3/4}$
\citep[\eg,][]{Zimmerman:05}. The observed `photosphere' or surface
of the disk has an effective temperature $T_{\rm eff}^{4} \sim
T^{4}/\tau$, where $\tau=\kappa \Sigma$ and $\kappa$ is the opacity
parameter.  In order for the observed flux to drop dramatically at
$\leq 3500$~\AA, the disk luminosity at $R < 150\, r_{g}$ must drop
significantly.

% by $\sim 2$ orders of magnitude.  If some small UV flux persists,
% the thin disk luminosity at $R<150r_{g}$ can drop by an order of
% magnitude.

We can parameterize the relevant disk timescales for a black hole
of mass $M_{\rm BH}$ at $R \sim 150r_{g}$ as:
\begin{eqnarray}
t_{\rm orb} & \sim & 10\, {\rm day}  \left(\frac{M_{\rm BH}}{10^8 M_{\odot}}\right)  \left(\frac{R}{150r_{g}}\right)^{3/2} \\
% t_{\rm orb} & \sim & 6\, {\rm week}  \left(\frac{M}{4 \times 10^{8}M_{\odot}}\right)  \left(\frac{R}{150r_{g}}\right)^{3/2} \\
% t_{\rm orb} & \sim & 6\, {\rm week} \left(\frac{R}{150r_{g}}\right)^{3/2} \\
t_{\rm th} & \sim & 1\, {\rm yr} \left( \frac{\alpha}{0.03}
  \right)^{-1} \left(\frac{M_{\rm BH}}{10^8 M_{\odot}}\right) \left(\frac{R}{150r_{g}}\right)^{3/2} \\
% t_{\rm th} & \sim & 4\, {\rm yr} \left( \frac{\alpha}{0.03}
%   \right)^{-1} \left(\frac{R}{150r_{g}}\right)^{3/2} \\
t_{\rm front} & \sim & 20\, {\rm yr} \left( \frac{h/R}{0.05}
  \right)^{-1} \left( \frac{\alpha}{0.03} \right)^{-1}
  \left(\frac{M_{\rm BH}}{10^8 M_{\odot}}\right)
  \left(\frac{R}{150r_{g}}\right)^{3/2} \\
% t_{\rm front} & \sim & 80\, {\rm yr} \left( \frac{h/R}{0.05}
%   \right)^{-1} \left( \frac{\alpha}{0.03} \right)^{-1}
%   \left(\frac{R}{150r_{g}}\right)^{3/2} \\
t_{\nu} & \sim & 400\, {\rm yr} \left( \frac{h/R}{0.05} \right)^{-2}
  \left( \frac{\alpha}{0.03} \right)^{-1} \left(\frac{M_{\rm BH}}{10^8 M_{\odot}}\right) \left( \frac{R}{150r_{g}}
  \right)^{3/2} 
% t_{\nu} & \sim & 1600\, {\rm yr} \left( \frac{h/R}{0.05} \right)^{-2}
%   \left( \frac{\alpha}{0.03} \right)^{-1} \left( \frac{R}{150r_{g}}
%   \right)^{3/2}. 
\end{eqnarray}
\noindent Since \qso\ has a black hole mass of $M_{\rm BH} =
4 \times 10^{8}M_{\odot}$, these timescales range from $\sim 6$~weeks
to 1600~yr for that source.  Since we observe changes on timescales
of a few years in this source, any changes in the disk are unlikely
to be happening on the dynamical (\ie, orbital) timescale, which
is far too short.  Nor are they likely to be due to viscous effects
or the propagation of cooling/heating fronts in the inner disk,
which have characteristic timescales that are too long.  In particular,
the changing-look phenomenon might naturally have been assumed
associated with the inner accretion disk clearing out, analogous
to the disk truncation believed associated with state changes in
Galactic binaries \citep[\eg,][]{Done:07, Neilsen:11}.  However,
such disk truncation happens on the viscous timescale, which is
orders of magnitude longer than the transitions observed in
changing-look quasars.  This rough parameterization above instead
suggests that the {\em thermal} timescale most closely matches the
observed source variability.

One caveat is that a revision of some of these
parameters might make the heating/cooling front scenario plausible.
The standard \citet{Shakura:73} thin disk assumes that once matter
plunges across the innermost stable circular orbit (ISCO), which
defines the inner edge of the accretion disk, the matter has no
connection to the disk.  This is called the zero-torque condition,
and the disk remains thin all the way to the ISCO.  If, instead,
some connection is maintained between the plunging material and the
innermost accretion disk, perhaps due to magnetic fields, the
plunging material produces a torque in the inner accretion disk,
causing it to heat up and inflate.  Models and simulations both
show this is more likely than a zero-torque condition
\citep[\eg,][]{Agol:00}, meaning that $(h/R) \sim 0.2$ might be a
more plausible value than the value of 0.05 assumed in equation~(4),
at least, close to the ISCO.  Since this modification to the
disk structure is due to a boundary condition at the ISCO, even in
the inflated non-zero torque condition, the disk should be thin
well before $150 r_g$.  Second, while numerical simulations
\citep[\eg,][]{Hirose:09, Davis:10} tend to derive estimates for
$\alpha$ consistent the value assumed in equation~(4), \citet{King:07}
argue that observations favor a typical range of $\alpha \sim 0.1
- 0.4$.  They suggest that incomplete physics in the numerical
simulations, including treatment of the global structure of the
magnetic field, might cause the simulations to underestimate $\alpha$.
If we instead assume a more viscous disk with $\alpha \sim 0.3$,
and further consider a region slightly closer to the black hole,
$R \sim 100\, r_g$, then $t_{\rm front}$ drops by a factor of $\sim
80$, becoming $\sim 1$~yr for \qso.  Owing to its quadratic dependence
on $(h/R)$, the viscous timescale $t_\nu$ drops even more precipitously:
it falls by a factor of $\sim 300$, becoming $\sim 5.4$~yr. These
timescales are consistent with the evolution observed in \qso\ and
many of the other changing-look quasars in the literature.  Indeed,
in our companion paper, \citet{Ross:18}, we suggest that the observed
spectral evolution in another mid-IR selected changing-look quasar
is driven by a cooling front propagating outward, and then reflecting
back inward as a heating front.

\subsection{Causes of Changes in Inner Disk}
\label{sec:disk}

The discussion above shows that changes in the innermost disk
occurring on the thermal or cooling/heating front timescale seem
to be the most plausible explanation for the year-scale variability
observed in changing-look quasars.  We next consider potential
triggers of these changes, which broadly can be broken into two
classes: (i) disk instabilities, and (ii) local perturbations due
to objects or events locally or elsewhere in the disk.

\subsubsection{Disk Instabilities}

Disk instabilities can occur for a variety of reasons and on a range
of timescales \citep[\eg,][]{Lightman:74, Shakura:76, Lightman:76}.
A classic, short timescale instability is due to an instability in
($\Sigma,T_{\rm eff}$) parameter space due to large changes in
opacity from recombination as a result of small changes in $\Sigma$
\cite[\eg,][]{Lin:86}. In this case $\Sigma$ changes locally
so that the local disk lies on the unstable part of the ($\Sigma,
T_{\rm eff}$) parameter space S-curve. Once this happens, $T_{\rm
eff}$ can drop by an order of magnitude on approximately the thermal
timescale $t_{\rm th}$ \citep[or some small multiple
thereof;][]{Shakura:76}.  Since the mid-plane temperature $T$ is
approximately unchanged, but $T^{4}=\tau T_{\rm eff}^{4}$, $\tau$,
the optical depth to the mid-plane, can increase dramatically. 
Recent work shows that this thermal instability seems to operate
in simulations with turbulence driven by the magneto-rotational
instability \citep[MRI;][]{Jiang:14}, though perhaps not in simulations
that include realistic iron line opacities and sufficiently high
metal abundances \citep{Jiang:16}.

In principle, another type of instability might cause the torque
condition to change at the ISCO.  Consider, for example, a quasar
that starts in a non-zero torque state, and therefore with an
inflated inner disk \citep[\eg][]{Sirko:03}.  If some
magneto-hydrodynamical instability caused the torque to rapidly
decrease, then the inner disk would cool and deflate, thereby
creating a changing-look quasar.  \citet{Ross:18} present an
interesting source where our preferred explanation is exactly such
a scenario.

\subsubsection{Perturbations Due To Objects/Events}
\label{sec:emri}

Changes in the inner disk state can also occur due to the presence
of local perturbers, such as an extreme mass ratio inspiral (EMRI)
event, or more distant changes in the accretion flow.  A change in
the local value of $\dot{\Sigma}$, which promotes instabilities in
($\Sigma,T_{\rm eff}$) such as those described above, might occur
due to embedded supernovae in AGN disks \citep{McKernan:14,
McKernan:17}, or stalling of inmigrating objects \citep{Bellovary:16}.

A large population of stellar mass black holes, stellar remnants
and stars are expected in AGN disks \citep[\eg,][]{Syer:91,
Artymowicz:93, McKernan:12}.  Torques from gas in the disk causes
these secondary objects to migrate in the disk and a fraction of
the secondaries will end up on the central supermassive black holes
in an EMRI event. Once a secondary object ends up in the innermost
regions of the accretion disk, its mass can become comparable to,
or even dominate, the co-rotating disk mass. From equation~(4), a
stellar mass black hole of mass $\sim 10\, M_{\odot}$ could dominate
the innermost gas flow on a timescale of decades.  Therefore the
spectral output of the inner disk can change on the timescales of
the EMRI.  Specifically, initially migration torques on any embedded
object in a disk whose mass is significantly less than the mass of
the gas in the disk will cause the orbit to decay.  The timescale
of this orbital decay will range from tens of kyr to tens of Myr,
depending on multiple factors, such as the starting point of the
migration in the disk, the mass of the object, the disk surface
density, and the disk aspect ratio.  For all reasonable parameters,
the timescale will be significantly larger than typical changing-look
quasar timescales.  Even once the migrating object has lost enough
angular momentum from these migration torques and plunges to the
central supermassive black hole, the relevant time scale becomes
the freefall timescale (equation 2), which is an order of magnitude
smaller than the changing-look quasar timescale.  We are left with
the most likely cause of the changing-look quasar phenomenon being
some thermal or magnetic instability triggering a major change in
the innermost accretion disk that then propagates on either the
thermal or the heating/cooling front timescale.

% The gravitational wave decay timescale $t_{\rm GW} of an isolated
% stellar mass black hole of mass $M_2$ around a supermassive black
% hole of mass $M_{\rm SMBH}$ is given by \citet{Peters:64}:
% \begin{equation}
% t_{\rm GW} \approx \frac{5}{128} \frac{c^{5}}{G^{3}}
% \frac{a_{b}^{4}}{M_{b}^{2} \mu_{b}} (1-e_{b}^{2})^{7/2},
% \label{eq:t_gw}
% \end{equation}
% \noindent where $M_{b}=M_{\rm SMBH}+M_{2}$ is the binary mass,
% $\mu_{b}=M_{\rm SMBH} M_{2} / M_{b}$ is the binary reduced mass and
% $(a_{b},e_{b})$ are the initial binary semi-major axis and eccentricity.
% Re-parameterizing and assuming low eccentricity, we find
% \begin{equation}
% t_{\rm GW} \approx 60,000\, {\rm yr} \left( \frac{a_b}{10\, r_{g}}\right)^{4}
% \left( \frac{M_{\rm SMBH}}{10^8\, M_{\odot}}\right)^2 \left(\frac{M_2}{10\, M_\odot}
% \right)^{-1}.
% \end{equation}
% \noindent 

\section{Conclusions}

We have presented \qso, one of the first mid-IR selected changing
look quasars reported thus far.  The source was identified on the
basis of its extreme variability in in the \wise/\neowise\ data
stream, having faded by more than a factor of two at 3.4 and
4.6~$\mu$m between 2010 and 2014.  Optical surveys show comparable
or greater fading over a slightly longer temporal baseline, while
a comparison of archival \rosat\ data from 1995 to Target of
Opportunity \swift\ observations obtained in 2017 show the source
has faded by at least a factor of 10 in the low-energy X-ray band.
Motivated by this extreme fading, we obtained a second-epoch optical
spectrum \qso\ in early 2017 to compare with the archival SDSS
spectrum from 2006.  Over the intervening decade, the strong blue
continuum has collapsed in this source, and most of the broad Balmer
lines have disappeared; only broad H$\alpha$ remains visible, albeit
at a significantly weaker level.

We use this source as a touchstone to discuss physical models of
abrupt quasar fading, a subject that has received considerable
attention of late thanks to the growing ranks of wide-area ground-based
optical surveys \citep[\eg,][]{LaMassa:15, Macleod:16, Ruan:16a,
Ruan:16b, Runnoe:16, Gezari:17, Sheng:17, Yang:17, Assef:18, Ross:18}.
In particular, we emphasize the unique value of multi-epoch mid-IR
photometry to test and exclude models that attempt to ascribe large
changes in quasar luminosities to obscuration by an intervening
cloud.  The large, parsec-scale size of the mid-IR emitting region
is too large to be extincted by an intervening cloud on the timescales
probed thus far, and the long wavelength baseline of optical through
mid-IR data provide a long leverarm with which to test if the
observed variability is consistent with observed extinction laws.  

Thus, when strong mid-IR variability is observed, the strong
indication is that the changing-look phenomenon is not due to
obscuration, but is rather due to changes in the innermost regions
of the accretion disk, at distances $\simlt 150~r_g$ (\ie, $\simlt
600$~AU).  We consider the range of relevant disk timescales at
this distance, and show that the several week orbital, or dynamical
timescale is far shorter than the typical changing-look quasar
variability timescale of several years, while the millennium-long
viscous timescale, which is the timescale on which the accretion
disk can become truncated, is far too long.  Importantly, these
results show that the changing-look quasar phenomenon is physically
distinct from related phenomenon seen more locally.  Many changing-look
Seyfert galaxies, at lower luminosity, are clearly associated with
obscuration by an intervening cloud \citep[\eg, NGC~1365;][]{Risaliti:02,
Rivers:15}, while the state changes observed in Galactic binaries,
\ie, systems with an accretion disk around a stellar-mass black
hole or neutron star, appear associated with disk truncation
\citep[\eg, GRS~1915+105;][]{Neilsen:11}.

We are instead left with the relevant timescale being the thermal
timescale, or potentially the cooling/heating front timescale.  We
briefly discuss various physical phenomena that could cause abrupt
changes in the temperature structure of the disk, such as: (i) a
rapid change in the torque at the ISCO radius, (ii) thermal disk
instabilities where minor changes in the disk surface density can
cause major opacity changes, and thus major temperature changes
\citep[\eg][]{Lightman:74, Lin:86}, and (iii) perturbations caused
by an object or an event, such as an extreme mass ratio in-spiral
event.

In the coming years, the Zwicky Transient Facility (ZTF) and the
Large Synoptic Survey Telescope (LSST), will begin providing deeper
multi-wavelength optical photometry with several-day cadence over
large swaths of the sky.  Many new changing-look quasars will be
identified in these data streams.  This will provide the exciting
opportunity to find events while they are in process, rather than
archivally.  Unfortunately, these facilities will likely have minimal
(ZTF) to zero (LSST) overlap with mid-IR survey missions such as
\wise\ and \spitzer, but real-time discoveries will allow rapid
follow-up and monitoring with both X-ray observations and optical/near-IR
spectroscopy.  Such data will be essential for disentangling which
trigger or triggers cause the thermal changes observed in the inner
accretion disks of changing-look quasars.

\acknowledgements

We thank the anonymous referee for a timely and informative
report, which has improved our manuscript.  We also thank Javier
Garc{\'i}a for useful comments on the manuscript and Nikita Kamraj
for assisting with the 2017 January Palomar observations.  This
publication makes use of data products from the {\it Wide-field
Infrared Survey Explorer}, which is a joint project of the University
of California, Los Angeles, and the Jet Propulsion Laboratory/California
Institute of Technology, funded by the National Aeronautics and
Space Administration.  This publication makes use of data products
from the Near-Earth Object Wide-field Infrared Survey Explorer
(NEOWISE), which is a project of the Jet Propulsion Laboratory/California
Institute of Technology. NEOWISE is funded by the National Aeronautics
and Space Administration.  CRTS was supported by the NSF grants
AST-1313422, AST-1413600, and AST-1518308.  DS acknowledges support
from NASA through ADAP award 12-ADAP12-0109.  BM and KESF are
supported by NSF PAARE AST-1153335, and with to thank JPL and Caltech
for support during their sabbatical in early 2017.  MJG, SGD, and
AJD acknowledge partial support from the NASA grant 16-ADAP16-0232,
and NSF grants AST-1413600 and AST-1518308.
NPR acknowledges support from the STFC and the Ernest Rutherford
Fellowship.  AMM acknowledges support from NASA through ADAP award
NNH17AE75I.  RJA acknowledges support from FONDECYT grant number
1151408.  MB gratefully acknowledges financial support from NASA
Headquarters under the NASA Earth and Space Science Fellowship
Program, grant NNX14AQ07H, and the support from the Black Hole
Initiative at Harvard University, which is funded by a grant from
the John Templeton Foundation.  AD's research was supported in part
by the National Optical Astronomy Observatory (NOAO). NOAO is
operated by the Association of Universities for Research in Astronomy
(AURA), Inc.  under a cooperative agreement with the National Science
Foundation.  HDJ acknowledges support from the Basic Science Research
Program through the National Research Foundation of Korea (NRF),
funded by the Ministry of Education (NRF-2017R1A6A3A04005158).

\bibliographystyle{apj.bst}

\begin{thebibliography}{84}
\expandafter\ifx\csname natexlab\endcsname\relax\def\natexlab#1{#1}\fi

\bibitem[{Agol \& Krolik(2000)}]{Agol:00}
Agol, E. \& Krolik, J.~H. 2000, \apj, 528, 161

\bibitem[{Arcavi {et~al.}(2014)Arcavi, Gal-Yam, Sullivan, {et~al.}}]{Arcavi:14}
Arcavi, I., Gal-Yam, G., Sullivan, M., {et~al.} 2014, \apj, 793, 38

\bibitem[{Artymowicz {et~al.}(1993)Artymowicz, Lin, \& Wampler}]{Artymowicz:93}
Artymowicz, P., Lin, D.~N.~C., \& Wampler, E.~J. 1993, \apj, 409, 592

\bibitem[{Assef {et~al.}(2010)Assef, Kochanek, Brodwin, {et~al.}}]{Assef:10}
Assef, R.~J., Kochanek, C.~S., Brodwin, M., {et~al.} 2010, \apj, 713, 970

\bibitem[{Assef {et~al.}(2013)Assef, Stern, Kochanek, {et~al.}}]{Assef:13}
Assef, R.~J., Stern, D., Kochanek, C.~S., {et~al.} 2013, \apj, 772, 26

\bibitem[{Assef {et~al.}(2018)Assef, Stern, Noirot, Jun, Cutri, \&
  Eisenhardt}]{Assef:18}
Assef, R.~J., Stern, D., Noirot, G., Jun, H.~D., Cutri, R.~M., \& Eisenhardt,
  P.~R. 2018, \apjs, 234, 23

\bibitem[{Becker {et~al.}(1995)Becker, White, \& Helfand}]{Becker:95}
Becker, R.~H., White, R.~L., \& Helfand, D.~J. 1995, \apj, 450, 559

\bibitem[{Bellovary {et~al.}(2016)Bellovary, {Mac~Low}, McKernan, \&
  Ford}]{Bellovary:16}
Bellovary, J., {Mac~Low}, M., McKernan, B., \& Ford, K.~E.~S. 2016, \apj, 819,
  L17

\bibitem[{Blagorodnova {et~al.}(2017)Blagorodnova, Gezari, Hung,
  {et~al.}}]{Blagorodnova:17}
Blagorodnova, N., Gezari, S., Hung, T., {et~al.} 2017, \apj, 844, 46

\bibitem[{Boller {et~al.}(2016)Boller, Freyberg, Tr\"umper,
  {et~al.}}]{Boller:16}
Boller, T., Freyberg, M.~J., Tr\"umper, J., {et~al.} 2016, \aap, 588, 103

\bibitem[{Burrows {et~al.}(2005)Burrows, Hill, Nousek, {et~al.}}]{Burrows:05}
Burrows, D.~N., Hill, J.~E., Nousek, J.~A., {et~al.} 2005, SSRv, 120, 165

\bibitem[{Chambers {et~al.}(2016)Chambers, Magnier, Metcalfe,
  {et~al.}}]{Chambers:16}
Chambers, K.~C., Magnier, E.~A., Metcalfe, N., {et~al.} 2016,
  (arXiv:1612.05560)

\bibitem[{Czerny(2006)}]{Czerny:06}
Czerny, B. 2006, in ASP Conf. Ser., AGN Variability from X-Rays to Radio Waves,
  ed. I.~M. McHardy, B.~M. Peterson, \& S.~G. Sergeev, Vol. 360 (ASP), 265

\bibitem[{Davis {et~al.}(2010)Davis, Stone, \& Pessah}]{Davis:10}
Davis, S.~W., Stone, J.~M., \& Pessah, M.~E. 2010, \apj, 713, 52

\bibitem[{Done {et~al.}(2007)Done, Gierli{\'n}ski, \& Kubota}]{Done:07}
Done, C., Gierli{\'n}ski, M., \& Kubota, A. 2007, A\&ARv, 15, 1

\bibitem[{Drake {et~al.}(2009)Drake, Djorgovski, Mahabal, {et~al.}}]{Drake:09}
Drake, A.~J., Djorgovski, S.~G., Mahabal, A., {et~al.} 2009, \apj, 696, 870

\bibitem[{Fabian \& Terlevich(1996)}]{Fabian:96}
Fabian, A.~C. \& Terlevich, R. 1996, \mnras, 280, L5

\bibitem[{Fitzpatrick(1999)}]{Fitzpatrick:99}
Fitzpatrick, E.~L. 1999, \pasp, 111, 63

\bibitem[{Gezari {et~al.}(2017)Gezari, Hung, Cenko, {et~al.}}]{Gezari:17}
Gezari, S., Hung, T., Cenko, S.~B., {et~al.} 2017, \apj, 835, 144

\bibitem[{Goodrich(1989)}]{Goodrich:89}
Goodrich, R.~W. 1989, \apj, 340, 190

\bibitem[{Graham {et~al.}(2017)Graham, Djorgovski, Drake, {et~al.}}]{Graham:17}
Graham, M.~J., Djorgovski, S.~G., Drake, A.~J., {et~al.} 2017, \mnras, 470,
  4112

\bibitem[{Graham {et~al.}(2015{\natexlab{a}})Graham, Djorgovski, Stern,
  {et~al.}}]{Graham:15}
Graham, M.~J., Djorgovski, S.~G., Stern, D., {et~al.} 2015{\natexlab{a}}, \nat,
  518, 74

\bibitem[{Graham {et~al.}(2015{\natexlab{b}})Graham, Djorgovski, Stern,
  {et~al.}}]{Graham:15b}
---. 2015{\natexlab{b}}, \mnras, 453, 1562

\bibitem[{Guo {et~al.}(2016)Guo, Malkan, Gu, {et~al.}}]{Guo:16}
Guo, H., Malkan, M.~A., Gu, M., {et~al.} 2016, \apj, 826, 186

\bibitem[{Hameury {et~al.}(2009)Hameury, Viallet, \& Lasota}]{Hameury:09}
Hameury, J., Viallet, M., \& Lasota, J. 2009, \aap, 496, 413

\bibitem[{Hirose {et~al.}(2009)Hirose, Blaes, \& Krolik}]{Hirose:09}
Hirose, S., Blaes, O., \& Krolik, J.~H. 2009, \apj, 704, 781

\bibitem[{Husemann {et~al.}(2016)Husemann, Urrutia, Tremblay,
  {et~al.}}]{Husemann:16}
Husemann, B., Urrutia, T., Tremblay, G.~R., {et~al.} 2016, \aap, 593, L9

\bibitem[{Ichikawa \& Tazaki(2017)}]{Ichikawa:17}
Ichikawa, K. \& Tazaki, R. 2017, \apj, 844, 21

\bibitem[{Jiang {et~al.}(2014)Jiang, Davis, \& Stone}]{Jiang:14}
Jiang, Y., Davis, S.~W., \& Stone, J.~M. 2014, \apj, 796, 106

\bibitem[{Jiang {et~al.}(2016)Jiang, Davis, \& Stone}]{Jiang:16}
---. 2016, \apj, 827, 10

\bibitem[{Jun {et~al.}(2015{\natexlab{a}})Jun, Im, Lee, {et~al.}}]{Jun:15a}
Jun, H.~D., Im, M., Lee, H.~M., {et~al.} 2015{\natexlab{a}}, \apj, 806, 109

\bibitem[{Jun {et~al.}(2015{\natexlab{b}})Jun, Stern, Graham,
  {et~al.}}]{Jun:15b}
Jun, H.~D., Stern, D., Graham, M.~J., {et~al.} 2015{\natexlab{b}}, \apj, 814,
  L12

\bibitem[{Kalberla {et~al.}(2005)Kalberla, Burton, Hartmann,
  {et~al.}}]{Kalberla:05}
Kalberla, P.~M.~W., Burton, W.~B., Hartmann, D., {et~al.} 2005, \aap, 440, 775

\bibitem[{Kankare {et~al.}(2017)Kankare, Kotak, Mattila, {et~al.}}]{Kankare:17}
Kankare, E., Kotak, R., Mattila, S., {et~al.} 2017, Nature Astronomy, 1, 865

\bibitem[{King {et~al.}(2007)King, Pringle, \& Livio}]{King:07}
King, A.~R., Pringle, J.~E., \& Livio, M. 2007, \mnras, 376, 1740

\bibitem[{Klein {et~al.}(1994)Klein, McKee, \& Colella}]{Klein:94}
Klein, R.~I., McKee, C.~F., \& Colella, P. 1994, \apj, 420, 213

\bibitem[{LaMassa {et~al.}(2015)LaMassa, Cales, Moran, {et~al.}}]{LaMassa:15}
LaMassa, S.~M., Cales, S., Moran, E.~C., {et~al.} 2015, \apj, 800, 144

\bibitem[{Lang(2014)}]{Lang:14}
Lang, D. 2014, \aj, 147, 108

\bibitem[{Lang {et~al.}(2016)Lang, Hogg, \& Schlegel}]{Lang:16}
Lang, D., Hogg, D.~W., \& Schlegel, D.~J. 2016, \aj, 151, 36

\bibitem[{Lawrence {et~al.}(2016)Lawrence, Bruce, MacLeod,
  {et~al.}}]{Lawrence:16}
Lawrence, A., Bruce, A.~G., MacLeod, C., {et~al.} 2016, \mnras, 463, 296

\bibitem[{Lightman \& Eardley(1974)}]{Lightman:74}
Lightman, A.~P. \& Eardley, D.~M. 1974, \apj, 187, L1

\bibitem[{Lightman \& Shapiro(1976)}]{Lightman:76}
Lightman, A.~P. \& Shapiro, S.~L. 1976, \apj, 203, 701

\bibitem[{Lin \& Shields(1986)}]{Lin:86}
Lin, D.~N.~C. \& Shields, G.~A. 1986, \apj, 305, 28

\bibitem[{Liu {et~al.}(2015)Liu, Gezari, Heinis, {et~al.}}]{Liu:15}
Liu, T., Gezari, S., Heinis, S., {et~al.} 2015, \apj, 803, 16

\bibitem[{Macleod {et~al.}(2016)Macleod, Ross, Lawrence, {et~al.}}]{Macleod:16}
Macleod, C.~L., Ross, N.~P., Lawrence, A., {et~al.} 2016, \mnras, 457, 389

\bibitem[{Mainzer {et~al.}(2014)Mainzer, Bauer, Grav, {et~al.}}]{Mainzer:14}
Mainzer, A., Bauer, J., Grav, T., {et~al.} 2014, \apj, 784, 110

\bibitem[{Massey \& Gronwall(1990)}]{Massey:90}
Massey, P. \& Gronwall, C. 1990, \apj, 358, 344

\bibitem[{Matt {et~al.}(2003)Matt, Guainazzi, \& Maiolino}]{Matt:03}
Matt, G., Guainazzi, M., \& Maiolino, R. 2003, \mnras, 342, 422

\bibitem[{Matthews \& Sandage(1963)}]{Matthews:63}
Matthews, T.~A. \& Sandage, A.~R. 1963, \apj, 138, 30

\bibitem[{McElroy {et~al.}(2016)McElroy, Husemann, Croom,
  {et~al.}}]{McElroy:16}
McElroy, R.~E., Husemann, B., Croom, S.~M., {et~al.} 2016, \aap, 593, L8

\bibitem[{McKernan {et~al.}(2017)McKernan, Ford, Bellovary,
  {et~al.}}]{McKernan:17}
McKernan, B., Ford, K.~E.~S., Bellovary, J., {et~al.} 2017, \mnras, submitted
  (arXiv:1702.07818)

\bibitem[{McKernan {et~al.}(2014)McKernan, Ford, Kocsis, Lyra, \&
  Winter}]{McKernan:14}
McKernan, B., Ford, K.~E.~S., Kocsis, B., Lyra, W., \& Winter, L.~M. 2014,
  \mnras, 441, 900

\bibitem[{McKernan {et~al.}(2012)McKernan, Ford, Lyra, \& Perets}]{McKernan:12}
McKernan, B., Ford, K.~E.~S., Lyra, W., \& Perets, H.~B. 2012, \mnras, 425, 460

\bibitem[{Mor \& Netzer(2012)}]{Mor:12}
Mor, R. \& Netzer, H. 2012, \mnras, 420, 526

\bibitem[{Neilsen {et~al.}(2011)Neilsen, Remillard, \& Lee}]{Neilsen:11}
Neilsen, J., Remillard, R.~A., \& Lee, J.~C. 2011, \apj, 737, 69

\bibitem[{Rafiee {et~al.}(2016)Rafiee, Pirkola, Hall, Galati, Rogerson, \&
  Ameri}]{Rafiee:16}
Rafiee, A., Pirkola, P., Hall, P.~B., Galati, N., Rogerson, J., \& Ameri, A.
  2016, \mnras, 459, 2472

\bibitem[{Ricci {et~al.}(2016)Ricci, Bauer, Arevalo, {et~al.}}]{Ricci:16}
Ricci, C., Bauer, F.~E., Arevalo, P., {et~al.} 2016, \apj, 820, 5

\bibitem[{Risaliti {et~al.}(2002)Risaliti, Elvis, \& Nicastro}]{Risaliti:02}
Risaliti, G., Elvis, M., \& Nicastro, F. 2002, \apj, 571, 234

\bibitem[{Rivers {et~al.}(2015)Rivers, Risaliti, Walton, {et~al.}}]{Rivers:15}
Rivers, E., Risaliti, G., Walton, D.~J., {et~al.} 2015, \apj, 804, 107

\bibitem[{Roming {et~al.}(2005)Roming, Kennedy, Mason, {et~al.}}]{Roming:05}
Roming, P.~W.~A., Kennedy, T.~E., Mason, K.~O., {et~al.} 2005, SSRv, 120, 95

\bibitem[{Ross {et~al.}(2018)Ross, Ford, Graham, {et~al.}}]{Ross:18}
Ross, N.~P., Ford, K.~E., Graham, M.~J., {et~al.} 2018, \mnras, submitted

\bibitem[{Ruan {et~al.}(2016{\natexlab{a}})Ruan, Anderson, Cales,
  {et~al.}}]{Ruan:16a}
Ruan, J.~J., Anderson, S.~F., Cales, S.~L., {et~al.} 2016{\natexlab{a}}, \apj,
  826, 188

\bibitem[{Ruan {et~al.}(2016{\natexlab{b}})Ruan, Anderson, Green,
  {et~al.}}]{Ruan:16b}
Ruan, J.~J., Anderson, S.~F., Green, P.~J., {et~al.} 2016{\natexlab{b}}, \apj,
  825, 137

\bibitem[{Runnoe {et~al.}(2016)Runnoe, Cales, Ruan, {et~al.}}]{Runnoe:16}
Runnoe, J.~C., Cales, S., Ruan, J.~J., {et~al.} 2016, \mnras, 455, 1691

\bibitem[{Schlafly \& Finkbeiner(2011)}]{Schlafly:11}
Schlafly, E.~F. \& Finkbeiner, D. 2011, \apj, 737, 103

\bibitem[{Shakura \& Sunyaev(1973)}]{Shakura:73}
Shakura, N. \& Sunyaev, R.~A. 1973, \aap, 24, 337

\bibitem[{Shakura \& Sunyaev(1976)}]{Shakura:76}
---. 1976, \mnras, 175, 613

\bibitem[{Shappee {et~al.}(2014)Shappee, Prieto, Grupe, {et~al.}}]{Shappee:14}
Shappee, B.~J., Prieto, J.~L., Grupe, D., {et~al.} 2014, \apj, 788, 48

\bibitem[{Shen {et~al.}(2011)Shen, Richards, Strauss, {et~al.}}]{Shen:11}
Shen, Y., Richards, G.~T., Strauss, M.~A., {et~al.} 2011, \apjs, 194, 45

\bibitem[{Sheng {et~al.}(2017)Sheng, Wang, Jiang, Yang, Yan, Dou, \&
  Peng}]{Sheng:17}
Sheng, Z., Wang, T., Jiang, N., Yang, C., Yan, L., Dou, L., \& Peng, B. 2017,
  \apj, 846, 7

\bibitem[{Sirko \& Goodman(2003)}]{Sirko:03}
Sirko, E. \& Goodman, J. 2003, \mnras, 341, 501

\bibitem[{Stathakis \& Sadler(1991)}]{Stathakis:91}
Stathakis, R.~A. \& Sadler, E.~M. 1991, \mnras, 250, 786

\bibitem[{Stern {et~al.}(2012)Stern, Assef, Benford, {et~al.}}]{Stern:12}
Stern, D., Assef, R.~J., Benford, D.~J., {et~al.} 2012, \apj, 753, 30

\bibitem[{Stern {et~al.}(2005)Stern, Eisenhardt, Gorjian, {et~al.}}]{Stern:05}
Stern, D., Eisenhardt, P., Gorjian, V., {et~al.} 2005, \apj, 631, 163

\bibitem[{Stern {et~al.}(2017)Stern, Graham, Arav, {et~al.}}]{Stern:17}
Stern, D., Graham, M.~J., Arav, N., {et~al.} 2017, \apj, 839, 2

\bibitem[{Storchi-Bergmann {et~al.}(1995)Storchi-Bergmann, Eracleous, Livio,
  Wilson, Filippenko, \& Halpern}]{StorchiBergmann:95b}
Storchi-Bergmann, T., Eracleous, M., Livio, M., Wilson, A.~S., Filippenko,
  A.~V., \& Halpern, J.~P. 1995, \apj, 443, 617

\bibitem[{Syer {et~al.}(1991)Syer, Clarke, \& Rees}]{Syer:91}
Syer, D., Clarke, C., \& Rees, M.~J. 1991, \mnras, 250, 505

\bibitem[{Tohline \& Osterbrock(1976)}]{Tohline:76}
Tohline, J.~E. \& Osterbrock, D.~E. 1976, \apjl, 210, L117

\bibitem[{Urry \& Padovani(1995)}]{Urry:95}
Urry, C.~M. \& Padovani, P. 1995, \pasp, 107, 803

\bibitem[{{Van~Dyk} {et~al.}(1993){Van~Dyk}, Weiler, Sramek, \&
  Panagia}]{VanDyk:93}
{Van~Dyk}, S.~D., Weiler, K.~W., Sramek, R.~A., \& Panagia, N. 1993, \apj, 419,
  69

\bibitem[{Walton {et~al.}(2014)Walton, Risaliti, Harrison,
  {et~al.}}]{Walton:14}
Walton, D.~J., Risaliti, G., Harrison, F.~A., {et~al.} 2014, \apj, 788, 76

\bibitem[{Wright {et~al.}(2010)Wright, Eisenhardt, Mainzer,
  {et~al.}}]{Wright:10}
Wright, E.~L., Eisenhardt, P.~R.~M., Mainzer, A.~K., {et~al.} 2010, \aj, 140,
  1868

\bibitem[{Yang {et~al.}(2017)Yang, Wu, Fan, {et~al.}}]{Yang:17}
Yang, Q., Wu, X., Fan, X., {et~al.} 2017, \apj, submitted (arXiv:1711.08122)

\bibitem[{Zimmerman {et~al.}(2005)Zimmerman, Narayan, McClintock, \&
  Miller}]{Zimmerman:05}
Zimmerman, E.~R., Narayan, R., McClintock, J.~E., \& Miller, J.~M. 2005, \apj,
  618, 832

\end{thebibliography}

\smallskip
{\it Facilities:} \facility{CRTS}, \facility{NEOWISE}, \facility{NOAO/CTIO
Blanco (DECam)}, \facility{Palomar (DBSP)},  \facility{Pan-STARRS},
\facility{SDSS}, \facility{WISE}

\smallskip
\copyright 2018.  All rights reserved.

\clearpage
\end{document}